
\input epsf

\input harvmac.tex

\def\figin{\epsfcheck\figin}\def\figins{\epsfcheck\figins}
\def\epsfcheck{\ifx\epsfbox\UnDeFiNeD
\message{(NO epsf.tex, FIGURES WILL BE IGNORED)}
\gdef\figin##1{\vskip2in}\gdef\figins##1{\hskip.5in}
\else\message{(FIGURES WILL BE INCLUDED)}
\gdef\figin##1{##1}\gdef\figins##1{##1}\fi}
\def\DefWarn#1{}
\def\figinsert{\goodbreak\midinsert}
\def\ifig#1#2#3{\DefWarn#1\xdef#1{fig.~\the\figno}
\writedef{#1\leftbracket fig.\noexpand~\the\figno}
\figinsert\figin{\centerline{#3}}\medskip\centerline{\vbox
{\baselineskip12pt
\centerline{\footnotefont{\bf Fig.~\the\figno:} #2}}}
\bigskip\endinsert\global\advance\figno by1}


\def\CF{{\cal F}} 
\def\CH{{\cal H}} 
 
\def\CT{{\cal T}}
\def\sgn{{\, \rm sgn}} 


\lref\DasJevA{S. Das and A Jevicki, Mod. Phys. Lett. A5(1990) 1639.}
\lref\DemEtAlA{K. Demeterfi, A. Jevicki, and J. Rodrigues, Nucl. Phys.
B362(1991) 173.}
\lref\DemEtAlB{K. Demeterfi, A. Jevicki, and J. Rodrigues, Nucl. Phys.
B365(1991) 499.}
\lref\DemEtAlC{K. Demeterfi, A. Jevicki, and J. Rodrigues, Mod. Phys.
Lett. A6(1991) 3199.}
\lref\GroKleA{D. Gross and I. Klebanov, Nucl. Phys. B352(1991) 671.}
\lref\JevA{A. Jevicki, {\it ``Developments in 2D string theory'',}
Lectures presented at the Spring School in String Theory, Trieste 1993,
hep-th/9309115.}
\lref\JevYonA{A. Jevicki and T. Yoneya, Brown Preprint BROWN-HEP-904,
hep-th/9305109.}
\lref\KarSakA{D. Karabali and B. Sakita, Int. Jour. Mod. Phys.
A6(1991) 5079.}
\lref\KleA{I. Klebanov, {\it ``String theory in two dimensions'',}
in ``String Theory and Quantum Gravity'', Proceedings of the Trieste
Spring School 1991, eds. J. Harvey et al., (World Scientific, Singapore,
1992).}
\lref\ManEtAlA{G. Mandal, A. Sengupta, and S. Wadia, Mod. Phys. Lett.
A6(1991) 1465.}
\lref\MooA{G. Moore, Nucl. Phys. B368(1992) 557.}
\lref\MooEtAlA{G. Moore, R. Plesser, and S. Ramgoolam, Nucl. Phys.
B377(1992) 143.}
\lref\MooPleA{G. Moore and R. Plesser, Phys. Rev. D46(1992) 1730.}
\lref\PolA{J. Polchinski, Nucl. Phys. B362(1991) 125.}
\lref\SenWadA{A. Sengupta and S. Wadia, Int. Jour. Mod. Phys.
A6(1991) 1961.}
\lref\DKR{K. Demeterfi, I. Klebanov, and J. Rodrigues, Princeton 
preprint PUPT-1416,\break
hep-th/9308036, to appear in Physical Review Letters.}


\Title{\vbox{\baselineskip12pt\hbox{PUPT-1425}\hbox{hep-th/9310075}}}
{\vbox{\centerline{Quantum Corrections in Collective Field Theory}}}
\centerline{{Edward Hsu}
\footnote{$\!\!^\dagger$}{E-mail address: \tt{hsu@puhep1.princeton.edu}}
and Igor Klebanov}
\vskip2pt\centerline{\it{Department of Physics}}
\vskip1pt\centerline{\it{Princeton University}}
\vskip1pt\centerline{\it{Princeton, NJ 08544 USA}}
\rm
\vskip .5in

\noindent
We review and extend the computation of scattering amplitudes of
tachyons in the $c=1$ matrix model using a manifestly finite
prescription for the collective field hamiltonian.
We give further arguments for the exactness of the cubic hamiltonian by
demonstrating the equality of the loop corrections in the collective field
theory with those calculated in the fermionic picture.

\Date{October 1993}


\newsec{Introduction}

In the past several years, matrix models have revolutionized
the study of two-dimensional string theory and quantum gravity.  
(For a recent review, see \refs{\JevA}).
By mapping the $c=1$ string to a theory of non-relativistic free
fermions, the one dimensional hermitian matrix model has allowed the
calculation of the free energy and correlation functions to all orders
in the loop expansion.
A very enlightening description of 2D strings is the field theoretic
formulation of Das and Jevicki \refs{\DasJevA}. 
Using the bosonic collective field representation, they derived
a cubic hamiltonian, which upon normal ordering, receives an
additional contribution from a linear tadople term.
This field theory reproduces the tree and loop
diagrams of tachyon scattering, the tachyon field being related to the
collective field via a non-local field redefinition which accounts for
the external leg factors one finds in the tachyon scattering
amplitudes, i.e.
\eqn\extleg{
\left<\prod_{i=1}^N \CT(q_i)\right> = 
\left(\prod_{i=1}^N -\mu^{|q_i|/2} {\Gamma(-|q_i|)\over
\Gamma(|q_i|)}\right)
\,A(q_1,\cdots ,q_N),
}
where $A$ is the Euclidean continuation of collective field $S$--matrix
element.

Although much work has gone into 
rederiving the results of the fermionic
(and Liouville) methods using collective fields, there remains some
doubt as to whether this simple cubic hamiltonian applies without
modification at higher loops \refs{\KarSakA}.
Indeed to date, the only calculation of
scattering amplitudes at loop level has been the two-point function.
Additionally, the integrals in the calculations of
\refs{\DemEtAlA,\DemEtAlB}
are unpleasant because 
of divergences which must be carefully regularized.
In this paper, we follow the approach proposed in \refs{\KleA},
which suggests that we take the double scaling limit of the matrix model
with $\mu <0$, or equivalently, with $\mu >0$ but taking as the natural
coordinate the conjugate momentum
$p$ of the classical coordinate $\lambda$ \refs{\ManEtAlA}.
The theory thus defined is manifestly finite.  
We rederive many of the results of
\refs{\DemEtAlA,\DemEtAlB} in this finite formalism, and we also
include a new result, the calculation of
the three-point function at one-loop. Our results further validate the
claim that the cubic hamiltonian is not renormalized,
and that the bosonization procedure is
finite and exact to all orders of perturbation theory.

This paper is organized as follows:
In \S2 we briefly review the derivation of the collective hamiltonian.  
We show that a double-scaled theory with negative $\mu$ eliminates
the spurious divergences.
Using this manifestly finite formalism, we
calculate the four-point and five-point tree amplitudes in \S3.
We verify the
agreement with matrix model results by Euclidean continuation 
and inclusion of the external leg factors \extleg.
\S4 is devoted to loop corrections:
We compute the two- and three-point functions at one loop.
Agreement with matrix model results is again confirmed.
We conclude in \S5 with some remarks about the future applications of the
bosonic calculations.
In the appendix we
list for reference some useful integrals needed in the computations.

\newsec{The Collective Field Approach}

We now review the derivation of the collective field hamiltonian using
the method of Gross and Klebanov \refs{\GroKleA}.
Consider the second quantized hamiltonian for a system of free
fermions with Planck constant $1/\beta$,
\eqn\fermham{
\hat h = \int d\lambda\,\Bigl\{{1\over 2\beta
^2}{\partial\Psi^\dagger\over\partial\lambda}
{\partial\Psi\over\partial\lambda}+U(\lambda)\Psi^\dagger\Psi-
\mu_F(\Psi^\dagger\Psi-N)\Bigr\}.
}
Introducing chiral fermions $\Psi_L$ and
$\Psi_R$, \fermham\ can be shown to be composed of 
chiral parts, $\CH = 2\beta\hat h = \CH_L + \CH_R$, 
i.e. the left and right movers do not mix.
The mixing of chiralities occurs only through the boundary conditions.
Upon bosonizing the fermion fields, we find, as
$\beta\rightarrow\infty$,
\eqn\bosham{
{:\CH :}={1\over 2}{\int_0^\infty\!\! d\tau} :
\left[ P^2+(X')^2 -{\sqrt\pi\over\beta v^2} \left(PX'P+
{1\over 3}(X')^3 \right) -{1\over 2\beta\sqrt\pi} X'
\left( {v''\over 3v^3} -{(v')^2\over 2v^4} \right) \right] :  \ ,
}
where $v(\lambda)$ is the velocity of the classical trajectory of a
particle at the Fermi level in the potential $U(\lambda)$,
\eqn\v{
v(\lambda)={d\lambda\over d\tau}=\sqrt{2(\mu_F-U(\lambda))}.
}
In the double scaling limit, all that survive are this cubic
interaction
and linear tadpole, both of order $g_{st} = 1/(\beta\mu)$,
where $\mu$ is defined as $\mu_c - \mu_F$, $\mu_c$ being the height of
the potential barrier $U$.
The equivalence of \bosham\ with the Das-Jevicki hamiltonian
was demonstrated in \refs{\GroKleA}.

In the usual double-scaled theory, $v(\tau) = \sqrt{2\mu}\sinh\tau$
near the quadratic maximum,
and \bosham\ diverges at the turning point $\tau = 0$.  One can
either carefully
regularize the theory near the turning point
\refs{\DemEtAlA,\DemEtAlB,\SenWadA} or
approach the double scaling limit with $\mu < 0$ \refs{\KleA,\ManEtAlA}.
In this
paper we shall utilize the latter method, which renders the entire
perturbative expansion manifestly finite.  
In \refs{\KleA,\ManEtAlA}, it is shown that positive and negative $\mu$ are
related by a simple interchange of the classical coordinate
$\lambda$ with its conjugate momentum $p$.  This interpretation allows
us to apply this method of extracting finite hamiltonians to any 
potential, in particular, to the deformed matrix model of 
\refs{\JevYonA}.
For negative $\mu$, there is no turning point, and $v(\tau)
= \sqrt{2|\mu|}\cosh\tau$ near the quadratic maximum.
Equation \bosham\ becomes
\eqn\negmuham{
{:\CH :}={1\over 2}{\int_{-\infty}^\infty\!\!\!\! d\tau}\!\! :
\!\!\left[\! P^2\!+\!(X')^2\! -\!{\sqrt\pi\over 2\beta |\mu| \cosh^2\tau}
\left(\!PX'P\!+\!{1\over 3}(X')^3\! \right)
\!-\!{1-{3\over 2}\tanh^2\tau\over 12\beta |\mu|\sqrt\pi\cosh^2\tau}X'
\!\right]\!\! :,
}
and the divergence at $\tau=0$ has disappeared.

Since the above hamiltonian does not mix
the chiralities, we will consider 
only the scattering of right movers described by
\eqn\hamright{
{:\!\CH_R\!:}\!=\!{1\over 4}{\int_{-\infty}^\infty\!\!\!\! d\tau}\! :
\!\left[(P-X')^2\!+\!{\sqrt\pi\over 6\beta |\mu| \cosh^2\tau}
(P-X')^3\!+\!{1-{3\over 2}\tanh^2\tau\over 12\beta |\mu|\sqrt\pi\cosh^2\tau}
(P-X')\right]\! :.
}
We follow the methods of \refs{\DemEtAlA,\DemEtAlB} in calculating 
scattering amplitudes,
using the hamiltonian formalism to evaluate
Feynman diagrams.
We will use the canonical oscillator basis
\eqn\canosc{
\eqalign{
X(t,\tau)&=\int_{-\infty}^\infty{dk\over\sqrt{4\pi |k|}}
\left(a(k)e^{i(k\tau-|k|t)}+a^\dagger(k)e^{-i(k\tau-|k|t)}\right),\cr
P(t,\tau)&=-i\int_{-\infty}^\infty{dk\over\sqrt{4\pi |k|}}
|k|\left(a(k)e^{i(k\tau-|k|t)}-
a^\dagger(k)e^{-i(k\tau-|k|t)}\right),
}}
with $[a(k),a^\dagger(k')] = \delta(k-k')$.  Inserting this into 
\hamright, $\CH_R$ assumes the form
$\CH_R = \CH_2 + \CH_3+ \CH_1$, with
\eqn\hamiltonains{
\eqalign{
\CH_2 &=\int_0^\infty\!\! dk\,a^\dagger(k)\,a(k),\cr
\CH_3 &={i\over 24\pi\beta |\mu|}\int_0^\infty\!\! dk_1 dk_2 dk_3
\,{\sqrt{k_1 k_2 k_3}}\,\bigl[
f(k_1+k_2+k_3)\,a(k_1)\,a(k_2)\,a(k_3)\,-\cr
&\qquad\qquad\quad 3f(k_1+k_2-k_3):a(k_1)\,a(k_2)\,
a^\dagger (k_3):\bigr] + {\rm h.c.},\cr
\CH_1 &=-{i\over 48\pi\beta |\mu|}\int_0^\infty\!\!
dk \,{\sqrt k}\, g(k) \left(a(k)-a^\dagger (k)\right),
}}
where
\eqn\fg{
\eqalign{
f(k)=\int_{-\infty}^\infty\!\! d\tau\,
{1\over \cosh^2\tau}e^{ik\tau}&=
{\pi\,k\over\sinh(\pi k/2)},\cr
g(k)=\int_{-\infty}^\infty\!\! d\tau\,
{1-{3\over 2}\tanh^2\tau\over
\cosh^2\tau}e^{ik\tau}&=
{\pi\,(k^3+2k)\over 4\sinh(\pi k/2)}.
}}
For our purposes of perturbatively calculating scattering amplitudes,
we may either use old fashioned time-ordered diagrams, or use
the Feynman rules defined by the cubic interaction $\CH_3$ and the
linear tadpole $\CH_1$ (relevant at the loop level)
with the propagators \refs{\DemEtAlA}
\eqn\propagators{
\eqalign{
\left< T(a(k_1,t) a^\dagger(k_2,0))\right> &= \delta(k_1-k_2)
\int{dE\over 2\pi}{i\over E-k_1+i\epsilon}e^{-iEt},\cr
\left< T(a^\dagger(k_1,t) a(k_2,0))\right> &= \delta(k_1-k_2)
\int{dE\over 2\pi}{-i\over E+k_1-i\epsilon}e^{-iEt}.
}}

\newsec{Review of Tree Calculations}

As a warm up, let us evaluate the tree level $S$--matrix,
\eqn\s{
S = 1-2\pi i\delta(E_i - E_f)T.
}
Various authors have used the collective field in deriving the exact
$S$--matrix
\refs{
\DemEtAlA\DemEtAlB{--}\KleA,\GroKleA\SenWadA\PolA\DemEtAlC{--}\MooPleA}.
In this section, we will only calculate the four-point and five-point
functions.
Each right-moving massless particle has energy equal to momentum, $E=k>0$,
and in what follows, we shall use them interchangeably.

Let us consider scattering of two
particles of momenta $k_1$ and $k_2$ into particles of
momenta $k_3$ and $k_4$.
{}From second order perturbation theory, we easily find for the
$s$-channel
\eqn\tfours{
T^{(s)} = {g_{st}^2\over 16\pi^2} k_1 k_2 k_3 k_4
\int_{-\infty}^\infty\!\! dk\,{k\,f^2(k_1+k_2-k)\over k_1+k_2-k+
i\epsilon \sgn(k)}.
}
To evaluate the integral in \tfours, we use
\eqn\prinval{
{1\over x\pm i\epsilon} = {\cal P}{1\over x}\mp i\pi\delta(x)
}
and the integral listed in the appendix. We find
\eqn\tfourS{
T^{(s)} = -{g_{st}^2\over 16\pi^2} k_1 k_2 k_3 k_4
\biggl({8\pi\over 3}+4\pi i|k_1+k_2|\biggr).
}
Likewise, one may evaluate the contribution from the
$t$- and $u$-channels, which have similar forms.
Summing over all three channels, we obtain for the total amplitude
\eqn\sfour{
S(k_1,k_2;k_3,k_4) = -{g_{st}^2\over 2}\delta(k_1+k_2-k_3-k_4)
\prod_{j=1}^4 k_j\,(|k_1+k_2|+|k_1-k_3|+|k_1-k_4|-2i).
}
Under Euclidean continuation $k_i\rightarrow i|q_i|$, where
$q_i>0$ for an incoming particle and $q_i<0$ for an outgoing particle. 
Upon inclusion of the external leg factors,
the Euclidean result agrees with the fermion calculations 
\refs{\MooA, \MooEtAlA}.

Next we investigate the five-point function.  Consider the
scattering of four particles of momenta $k_1$, $k_2$, $k_4$, and
$k_5$ into one with momentum $k_3$.
The basic building block is
\eqn\tfivea{
{i{g_{st}^3}\over 64\pi^3}\int_{-\infty}^\infty\!\! dp_1\,dp_2\,
{p_1\,p_2\,f(k_1+k_2-p_1)\,f(k_3-p_1+p_2)\,f(k_4+k_5+p_2) \over
(k_1+k_2-p_1+i\epsilon \sgn(p_1))\,(k_4+k_5+p_2-i\epsilon \sgn(p_2))}
}
where $p_1$ and $p_2$ are the internal momenta.  Again we utilize the
integrals collected in the appendix and find that \tfivea\ yields
\eqn\tfiveA{
{i{g_{st}^3}\over 8\pi}\biggl(
{4\over 3}(k_1+k_2)(k_4+k_5)
-{2i\over 3}(k_1+k_2+k_4+k_5)
-{8\over 15}\biggr)
}
As in the four-point case, we must sum over the inequivalent
permutations of the momenta.  It
is shown in \refs{\DemEtAlB} that when we restrict the kinematic region,
we obtain exact agreement with the 
calculations in Liouville theory.

\newsec{One Loop Corrections}

{In order to determine the exactness of the bosonization procedure in \S2,
we study the quantum corrections to the scattering amplitudes
with the hope that \hamright\ is sufficient to reproduce all the
results of the fermionic theory.
Let us first look at the\parfillskip=0pt\par}
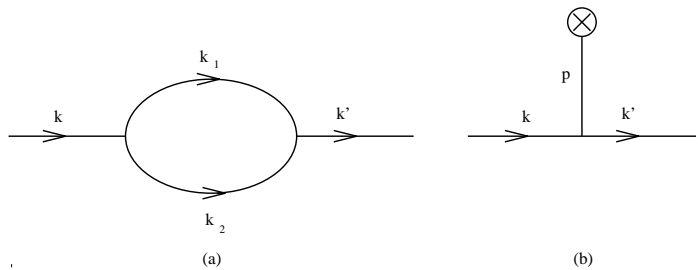
\ifig\twoone{The contributions to the two-point function of order
$g_{st}^2$.}{\epsfxsize3.0in\epsfbox{fig1.eps}}
\noindent two-point
function at one-loop given by the
diagrams in \twoone.  The
contribution to $T$ from the one-loop graph in \twoone $a$ is
\eqn\ttwoonea{
T_a = {g_{st}^2 k^2\over 32\pi^2}\int_0^\infty\!\! dk_1 dk_2\,
k_1 k_2 \left({f^2(k_1+k_2-k)\over k-k_1-k_2+i\epsilon}
-{f^2(k_1+k_2+k)\over k+k_1+k_2-i\epsilon}\right),
}
where the two terms come from the two different time orderings of the
two vertices.  One may of course also derive \ttwoonea\ using the Feynman
rules \propagators.  In that case, one needs to perform an integral over
the energy, which is conserved at the vertices, unlike the momentum.
After changing variables to $s=k_1+k_2$ and $k_2$, and
integrating over $k_2$, this becomes
\eqn\ttwooneA{
T_a = {g_{st}^2 k^2\over 192\pi^2}\int_{-\infty}^\infty\!\! ds\,
s^3 {f^2(k-s)\over k-s+i\epsilon \sgn(s)} = -{g_{st}^2 k^2\over
48\pi}\biggl(ik^3 + 2k^2+{8\over15}\biggr)
}
where the integral is evaluated using \prinval\ and integrals shown in
the appendix.

The contribution to $T$ from the tadpole graph \twoone $b$ is 
\eqn\ttwooneB{
T_b = {g_{st}^2 k^2\over 192\pi^2} \int_{-\infty}^\infty\!\! dp\,
{p\,f(p)\,g(p)\over -p+i\epsilon \sgn(p)}
= -{g_{st}^2 k^2\over 48\pi}\biggl({7\over15}\biggr)
}
This added to \ttwooneA\ gives
\eqn\ttwoone{
T = -{g_{st}^2 \over 48\pi}k^2(ik^3+2k^2+1).
}
By continuing to Euclidean space and including the external leg factors,
we again obtain complete agreement with the fermionic result
\refs{\MooA, \MooEtAlA}.

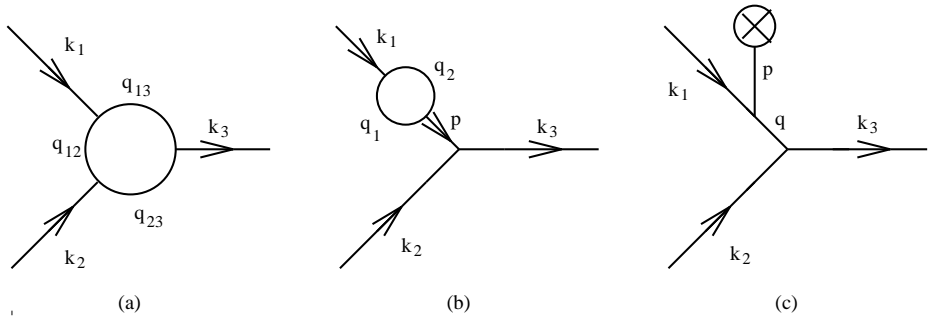
\ifig\threeone{The contributions to the three-point function of order 
${g_{st}}^2$.}
{\epsfxsize4.0in\epsfbox{fig2.eps}}

We now turn to the more difficult calculation of the $2\to 1$ amplitude
at one loop, where $k_1$ and $k_2$ are the incoming momenta.
This is another non-trivial check on the exactness of the hamiltonian
\bosham.
There are three types of diagrams, shown in \threeone. 
For \threeone$a$, we have six integrals, corresponding to the 
six possible time orderings for the three
vertices. Thus
\eqn\tthreeonea{
\eqalign{
T_a = k_1 &k_2 k_3 \Bigl({-i {g_{st}}^3 \over 64\pi^3}\Bigr)
\biggl\{ \cr
&{\int_0^\infty\!\! dq_{12} \int_0^\infty\!\! dq_{13}
\int_0^\infty\!\! dq_{23} {\CF\over (k_1-q_{12}-q_{13}+i\epsilon)
(k_3-q_{13}-q_{23}+i\epsilon)}}\cr
-&{\int_{-\infty}^0\!\! dq_{12} \int_0^\infty\!\! dq_{13}
\int_0^\infty\!\! dq_{23}
{\CF\over (k_2+q_{12}-q_{23}+i\epsilon)
(k_3-q_{13}-q_{23}+i\epsilon)}}\cr
+&{\int_0^\infty\!\! dq_{12} \int_0^\infty\!\! dq_{13}
\int_{-\infty}^0\!\! dq_{23} {\CF\over (k_1-q_{12}-q_{13}+i\epsilon)
(k_2+q_{12}-q_{23}-i\epsilon)}}\cr
-&{\int_{-\infty}^0\!\! dq_{12} \int_{-\infty}^0\!\! dq_{13}
\int_0^\infty\!\! dq_{23}
{\CF\over (k_1-q_{12}-q_{13}-i\epsilon)
(k_2+q_{12}-q_{23}+i\epsilon)}}\cr
+&{\int_0^\infty\!\! dq_{12} \int_{-\infty}^0\!\! dq_{13}
\int_{-\infty}^0\!\! dq_{23}
{\CF\over (k_2+q_{12}-q_{23}-i\epsilon)
(k_3-q_{13}-q_{23}-i\epsilon)}}\cr
-&{\int_{-\infty}^0\!\! dq_{12} \int_{-\infty}^0\!\! dq_{13}
\int_{-\infty}^0\!\! dq_{23}
{\CF\over (k_1-q_{12}-q_{13}-i\epsilon)
(k_3-q_{13}-q_{23}-i\epsilon)}}\biggr\},
}}
where $\CF = q_{12}\,q_{13}\,
q_{23}\,f(k_1-q_{12}-q_{13})\,f(k_2+q_{12}-q_{13})
\,f(k_3-q_{13}-q_{23})$.
The difficulty in evaluating \tthreeonea\ lies in the fact that some of
the limits of integration are not infinite, and under a change of 
variables such as $x=k_1-q_{12}-q_{13}$,
they aquire a finite $k$ dependence.  Under such
circumstances, it would be difficult to compute the integrals in 
\tthreeonea.
In order to circumvent this problem, we use the following identity
\eqn\propident{
\eqalign{
{1\over (k_1-q_{12}-q_{13}\pm i\epsilon)\,
(k_2+q_{12}-q_{23}\mp i\epsilon)}=
\qquad & \qquad\cr
{1\over(k_3-q_{13}-q_{23})}
\Bigl({1\over k_1-q_{12}-q_{13}\pm i\epsilon}
&+{1\over k_2+q_{12}-q_{23}\mp i\epsilon}\Bigr).
}}
This allows the third and fourth integrals in \tthreeonea\ to be split
into four, and upon combining some regions of integration,
equation \tthreeonea\ is transformed to
\eqn\tthreeonecomb{
\eqalign{
T_a =&k_1 k_2 k_3 \Bigl({-i {g_{st}}^3 \over 64\pi^3}\Bigr) \biggl\{ \cr
&\biggl(\int_0^\infty\!\!\!\! dq_{12}
\int_0^\infty\!\!\!\! dq_{13} \int_{-\infty}^
\infty\!\!\!\! dq_{23}
-\int_{-\infty}^0\!\!\!\! dq_{12} \int_{-\infty}^0\!\!\!\! dq_{13}
\int_{-\infty}^ \infty\!\!\!\! dq_{23}\biggr)
{\CF/(k_3-q_{13}-q_{23})\over k_1-q_{12}-q_{13}+i\epsilon \sgn(q_{12})}\cr
+&\biggl(\int_0^\infty\!\!\!\! dq_{12} \int_{-\infty}^\infty\!\!\!\! dq_{13}
\int_{-\infty}^0\!\!\!\! dq_{23}
-\int_{-\infty}^0\!\!\!\! dq_{12} \int_{-\infty}^\infty\!\!\!\! dq_{13}
\int_0^\infty\!\!\!\! dq_{23}\biggr)
{\CF/(k_3-q_{13}-q_{23})\over k_2+q_{12}-q_{23}-i\epsilon \sgn(q_{12})}\cr
-i\pi&\biggl(\int_0^\infty\!\!\!\! dq_{12}
\int_0^\infty\!\!\!\! dq_{13} \int_0^\infty\!\!\!\! dq_{23}
+\int_{-\infty}^0\!\!\!\! dq_{12} \int_{-\infty}^0\!\!\!\! dq_{13}
\int_{-\infty}^0\!\!\!\! dq_{23}\biggr)
{\CF\,\delta(k_3-q_{13}-q_{23})\over k_1-q_{12}-q_{13}+i\epsilon
\sgn(q_{12})}\cr
+i\pi&\biggl(\int_0^\infty\!\!\!\! dq_{12} \int_{-\infty}^0\!\!\!\! dq_{13}
\int_{-\infty}^0\!\!\!\! dq_{23}
+\int_{-\infty}^0\!\!\!\! dq_{12} \int_0^\infty\!\!\!\! dq_{13}
\int_0^\infty\!\!\!\! dq_{23}\biggr)
{\CF\,\delta(k_3-q_{13}-q_{23})\over k_2+q_{12}-q_{23}-i\epsilon
\sgn(q_{12})}\biggr\}.
}}
We now follow the same procedure as in the one-loop two-point function:
Make the change of variables to $s = q_{12}+q_{13}$,
$t=q_{23}-q_{12}$, and $q_{12}$, and integrate over $q_{12}$.
Equation \tthreeonecomb\ reduces to 
\eqn\tthreeonefinal{
\eqalign{
T_a = k_1 & k_2 k_3 \Bigl({-i {g_{st}}^3 \over 768
\pi^3}\Bigr) \biggl\{ \cr
&\int_{-\infty}^\infty\!\!\!\! ds \int_{-\infty}^\infty\!\!\!\! dt\,
s^3\,(s+2t)
{f(k_1-s)\,f(k_2-s)\,f(k_3-s-t) \over 
(k_1-s+i\epsilon \sgn(s))(k_3-s-t+i\epsilon \sgn(s+t))}\cr
+&\int_{-\infty}^\infty\!\!\!\! ds \int_{-\infty}^\infty\!\!\!\! dt\,
t^3\,(t+2s)
{f(k_1-s)\,f(k_2-s)\,f(k_3-s-t) \over
(k_2-t+i\epsilon \sgn(t))(k_3-s-t+i\epsilon \sgn(s+t))}\biggr\}.
}}
It is now straightforward to evaluate these integrals, using \prinval\ 
and the integrals tabulated in the appendix.  We need only calculate the
first integral in \tthreeonefinal\ since the second is the same
with $k_1$ and $k_2$ interchanged.
The result is  
\eqn\tthreeoneA{
T_a\!\!=\!\!{-i {g_{st}}^3 \over 48\pi}k_1 k_2 k_3\biggl(
\!-{1\over3}\bigl(k_1^4\!+\!2k_1^3 k_2\!+\!2k_1 k_2^3
\!+\! k_2^4\bigr) \!+\!
{2i\over3}\bigl(k_1^3\!+\!k_2^3\!+\!k_3^3\bigr) \!+\!
{8\over5}\bigl(k_1^2 \!+\! k_1 k_2 \!+\! k_2^2\bigr) \!+\!
{16\over35}\biggr).
}

Next, we look at \threeone$b$, where we have a loop on the external leg
$k_1$.  After making the substitution $s = q_1 + q_2$ and integrating
over $q_2$, we have
\eqn\tthreeonebone{
T_{b_1} = k_1 k_2 k_3 \Bigl({-i {g_{st}}^3 \over 768\pi^3}\Bigr)
\int_{-\infty}^\infty\!\!\!\! ds \int_{-\infty}^\infty\!\!\!\! dp\,p\,s^3
{f(k_1-p)\,f(k_1-s)\,f(p-s)\over(k_1-s+i\epsilon \sgn(s))
(k_1-p+i\epsilon \sgn(p))}.
}
Using the integrals in the appendix, this gives
\eqn\tthreeoneBone{
T_{b_1} = {-i {g_{st}}^3 \over 48\pi}k_1 k_2 k_3
\biggl(-{k_1^4\over3}+{4ik_1^3\over3}+{6k_1^2\over5}+
{4ik_1\over15}+{32\over105}\biggr).
}
We must, of course, 
also include the contributions from the diagrams where the loop
is attached to  $k_2$ and $k_3$, $T_{b_2}$ and $T_{b_3}$ respectively.
The total contribution of all three diagrams is
\eqn\tthreeoneB{
\eqalign{
T_b = &T_{b_1}+T_{b_2}+T_{b_3}\cr
=&{-i {g_{st}}^3 \over 48\pi}k_1 k_2 k_3
\biggl(-{1\over3}(k_1^4+k_2^4+k_3^4)
+{4i\over3}(k_1^3+k_2^3+k_3^3)\cr
&\qquad\qquad\qquad+{6\over5}(k_1^2+k_2^2+k_3^2)
+{4i\over15}(k_1+k_2+k_3)+{32\over35}
\biggr).
}}

Finally, there are the three diagrams where a tadpole
is attached on an external leg, as shown for the case of 
the tadpole on $k_1$ in \threeone $c$.  For the tadpole on $k_1$, we have
\eqn\tthreeoneconeCone{
\eqalign{
T_{c_1} = &k_1 k_2 k_3 \Bigl({-i {g_{st}}^3 \over768\pi^3}\Bigr)
\int_{-\infty}^\infty\!\!\!\! dp \int_{-\infty}^\infty\!\!\!\! dq\,
{p\,q\,g(p)\,f(k_1-q)\,f(k_1+p-q)\over(-p+i\epsilon \sgn(p))
(k_1-q+i\epsilon \sgn(q))}\cr
= &{-i {g_{st}}^3 \over 48\pi}k_1 k_2 k_3
\biggl({7ik_1\over30}+{22\over105}\biggr).
}}
Again we must include the diagrams with the tadpole attached to $k_2$
and $k_3$, which gives a total contribution of
\eqn\tthreeoneC{
T_c = {-i {g_{st}}^3 \over 48\pi}k_1 k_2 k_3
\biggl({7i\over30}(k_1+k_2+k_3)+{22\over35}\biggr).
}
By including the factor of  $-2\pi i\delta(k_1+k_2-k_3)$ and adding
equations \tthreeoneA , \tthreeoneB , and \tthreeoneC, we get
the total three-point function at one loop:
\eqn\sthreeone{
S(k_1, k_2;k_3) = -{g_{st}^3\over 24}\delta(k_1+k_2-k_3)k_1 k_2 k_3
\biggl((1+ik_3)(2+ik_3)(1+k_1^2+k_2^2-ik_3)\biggr).
}
Upon the Euclidean continuation $k_j\rightarrow i|q_j|$, and inclusion
of the external leg factors, this agrees with the 3-tachyon correlator
of the non-relativistic fermion calculation \refs{\MooA, \MooEtAlA}:
\eqn\sthreeoneferm{
\eqalign{
\langle \CT(q_1)&\CT(q_2)\CT(q_3)\rangle =\cr
&-{1\over 24 (\beta\mu)^3}
\delta (\sum_{j=1}^3 q_j) \prod_{j=1}^3\left(
{\Gamma(1-|q_j|)\over\Gamma(|q_j|)}\,\mu^{|q_j|/2}\right)
(|q_3|-1)(|q_3|-2)(q_1^2+q_2^2-|q_3|-1).
}}
The agreement of the bosonic calculations of collective field
theory with the fermionic results gives us confidence that
the bosonization procedure and 
the simple cubic hamiltonian are indeed
exact, to all orders in perturbation
theory.

\newsec{Conclusion}

In this paper we have reviewed computations of scattering amplitudes
using collective field theory in a manifestly finite formalism.
We have also presented some new calculations.
Furthermore, these results were shown to 
coincide with their fermionic counterparts, providing evidence that the
bosonization procedure
is finite and exact.

One drawback of the collective field theory
calculations is that the evaluation of higher point or
higher loop amplitudes become increasingly laborious.  As a purely
computational tool, the bosonized theory does not compare favorably with
its fermionic parent.  The methods of \refs{\MooEtAlA} are much more
powerful in their applications.
However, the bosonic
theory deserved study in its own right as a simple 
string field theory, where different 
backgrounds may be studied.  
Recently, Jevicki and Yoneya \refs{\JevYonA} proposed a deformed matrix
model and conjectured that it
describes the 2D black hole solution of critical string
theory.  
It is an interesting step 
towards understanding the $c=1$ theory in other backgrounds.
The results of this paper may be applied to that model,
and it would be interesting to show exactly
how the one-loop three-point
function vanishes, as found in \refs{\DKR}.

\bigskip
\centerline{\bf Acknowledgements}

We are indebted to K. Demeterfi for many useful comments.
We would also like to thank K. Ganga, S. Kachru,
D. Petrich, and M. Potters for helpful discussions.
This work was supported in part by DOE grant DE-AC02-76WRO3072,
the NSF Presidential Young Investigator Award PHY-9157482, 
James S. McDonnell Foundation grant No. 91-48,
and an A. P. Sloan Foundation Research Fellowship.

\vskip .2in

\appendix{A}{Some Useful Integrals}

In this appendix we list some of the integrals
necessary to evaluate the various
diagrams in this paper.
The first integral we encounter
occurs in the calculation of the 
four-point function
and many subsequent integrals:
\eqn\afour{
\int_{-\infty}^\infty\!\!\!\! dx\, f^2(x) = {8\pi\over3}.
}
For the five-point function integrals, we have
\eqn\afive{
\eqalign{
\int_{-\infty}^\infty\!\!\!\! dx \int_{-\infty}^\infty\!\!\!\! dy\,
{f(x)\,f(y)\,f(x-y) \over x\, y}
&={8\over 3}\pi^2,\cr
\int_{-\infty}^\infty\!\!\!\! dx \int_{-\infty}^\infty\!\!\!\! dy\,
f(x)\,f(y)\,f(x-y)
&={64\over 15}\pi^2,
}}
where in evaluating the integrals, we need to use the integral
definition of $f(x)$, equation \fg.

In addition to \afour, the following is needed for the loop diagram
\ttwooneA :
\eqn\atwoonea{
\int_{-\infty}^\infty\!\!\!\! dx\, x^2\,f^2(x) = {32\pi\over 15}.
}
For the tadpole integral in \ttwooneB, one finds
\eqn\atwooneb{
\int_{-\infty}^\infty\!\!\!\! dx\, f(x)\,g(x) = {28\pi\over 15}.
}
Below we list the integrals required for
the evaluation of diagram \threeone$a$ and \threeone$b$:
\eqn\athreeonea{
\eqalign{
\int_{-\infty}^\infty\!\!\!\! dx \int_{-\infty}^\infty\!\!\!\! dy\,
{x\, f(x)\,f(y)\,f(x-y)\over y}
&={32\over 15}\pi^2,\cr
\int_{-\infty}^\infty\!\!\!\! dx \int_{-\infty}^\infty\!\!\!\! dy\,
x^2\, f(x)\,f(y)\,f(x-y)
&={512\over 105}\pi^2,\cr
\int_{-\infty}^\infty\!\!\!\! dx \int_{-\infty}^\infty\!\!\!\! dy\,
{x^3\, f(x)\,f(y)\,f(x-y)\over y}
&={128\over 21}\pi^2.
}}
Finally, we have the tadpole integral \tthreeoneC, for which we must
have
\eqn\athreeonec{
\int_{-\infty}^\infty\!\!\!\! dx\int_{-\infty}^\infty\!\!\!\! dy\,
f(x)\,g(y)\,f(x+y) = {352\over 105}\pi^2.
}

\listrefs
\bye